# Non-Hermitian reshaping of high-order Landau modes

Zhihao Wang[1]†, Jie Jiang[2]†, Yanji Zheng[1], Wen Zhao[1], Chenyang Wang[3], Zhiwei Guo[2*], Yong-Chun Liu[3,4*], Shuang Zhang[5,6,7*], Cuicui Lu[1*]

[1]Key Laboratory of Advanced Optoelectronic Quantum Architecture and Measurements of Ministry of Education, State Key Laboratory of Chips and Systems for Advanced Light Field Display, Center for Interdisciplinary Science of Optical Quantum and NEMS Integration, Beijing Key Laboratory of Nanophotonics and Ultrafine Optoelectronic Systems, School of Physics, Beijing Institute of Technology, Beijing 100081, China.

[2]MOE Key Laboratory of Advanced Micro-Structured Materials, School of Physics Science and Engineering, Tongji University, Shanghai 200092, China.

[3]State Key Laboratory of Low-Dimensional Quantum Physics, Department of Physics, Tsinghua University, Beijing 100084, China.

[4]Frontier Science Center for Quantum Information, Beijing 100084, China.

[5]New Cornerstone Science Laboratory, Department of Physics, The University of Hong Kong, Hong Kong 999077, China.

[6]Quantum Science Center of Guangdong-Hong Kong-Macao Great Bay Area, Shenzhen 518045, China.

[7]Materials Innovation Institute for Life Sciences and Energy (MILES), The University of Hong Kong-Shenzhen Institute of Research and Innovation, Shenzhen 518057, China.

*Corresponding authors. Email: 2014guozhiwei@tongji.edu.cn(Z. G); ycliu@tsinghua.edu.cn (Y. L.); shuzhang@hku.hk (S. Z.); cuicuilu@bit.edu.cn (C. L.)

†These authors contributed equally to this work.

# Abstract

When charged particles are subjected to strong magnetic fields, they form discrete energy levels known as Landau levels. The Landau levels consist of a series of degenerate states of Landau modes, making them a promising platform for large-capacity information processing. However, to date, exploiting the high-order Landau modes and control their spatial distributions has remained elusive. Here, we propose to construct magnetic fields, electric fields, and imaginary momentum simultaneously to reshape high-order Landau modes in non-Hermitian systems. By building a non-Hermitian electric circuit platform, we experimentally realize pseudomagnetic fields via inhomogeneous coupling and pseudoelectric fields via a gradient on-site potential, while simultaneously introducing an imaginary momentum via non-reciprocal coupling. We directly observe multi-frequency single-peak localization of high-order Landau modes. Our work provides a universal method for manipulating high-order Landau modes and exploring applications in non-Hermitian systems, such as frequency multiplexing and wave packet reshaping.

# 1. Introduction

The behavior of charged particles in magnetic fields is the cornerstone of modern physics, underpinning phenomena such as Landau quantization and the quantum Hall effect. Landau levels are the foundation of the quantum Hall effects, where the Hall resistance becomes quantized and robust against disorder and irregularities in the sample [1]. This phenomenon is observed not only in conventional electron systems but also in massless Dirac systems with linear dispersion [2–4]. In recent years, it has been discovered that introducing strain or spatial inhomogeneities in Dirac systems can generate pseudo-magnetic field (PMF), also called artificial gauge field, which emulates the effects of a magnetic field without breaking time-reversal symmetry [5–7], enabling the generation of Landau levels and localized Landau modes in various systems [6,8–23]. With their vast degeneracy and inherent robustness against disorder and external perturbations, the Landau modes present a promising platform for large-capacity information processing. Thus, the study and utilization of Landau modes have become important and drawn extensive attention. For instance, recent work has realized Landau rainbow by breaking the translation symmetry of the system [13]. However, previous investigations have predominantly focused on the zero-order ($n = 0$) Landau modes. High-order Landau modes ($|n| = 1,2,\cdots$) exhibit rich multi-peak profiles, which provide more opportunities for mode reshaping. Moreover, high-order Landau modes provide a much broader operational bandwidth due to the infinite Landau level index. The absence of a universal analytical method hinders the manipulation of these high-order Landau modes.

In this work, we reshape Landau modes in non-Hermitian systems through the joint action of constructed magnetic fields, electric fields, and imaginary momentum. Specifically, the electric field breaks the degeneracy and separates the Landau modes in both energies and spatial positions. We analytically derive the degeneracy-lifted Landau modes and establish a direct mapping between their energy and spatial position. Concurrently, imaginary momentum is introduced to reshape the Landau mode envelopes and enhance their localization. As a result, distinct high-order Landau modes with different frequencies are separated and localized at various positions. As a concrete example, we show the reconfiguration of high-order Landau modes due to the modulation in non-Hermitian Dirac systems. Experimentally, we directly observe the frequency-dependent spatial localization of the high-order Landau modes on an electric circuit platform, where the PMF is generated by inhomogeneous coupling, the pseudo-electric field (PEF) is realized by a gradient on-site potential, and the imaginary momentum is introduced via non-reciprocal coupling. Our work not only opens avenues for modulating Landau modes, but also furnishes a topological platform for exploring applications in wave packet reshaping and frequency multiplexing.

## 2. Materials and Methods

Impedance spectra of the circuit were measured using a vector network analyzer (VNA, Tektronix TTR500) via preset ports between adjacent nodes. The N-port S-parameter matrix S was transformed into the admittance matrix (circuit Laplacian) using the relation $Y = (I - S)/(I + S)Z_0$, where $I$ is the identity matrix and $Z_0$ is the port impedance of the VNA. The complete matrix $Y$ contains all the information about the admittance eigenvalues and eigenstates. For steady-state voltage response measurements, AC current feeds, generated from a voltage source, were injected into the node through shunt resistances. The voltage distributions on all nodes were then recorded at different driving frequencies using an oscilloscope (Keysight DSOX4024A). Circuit simulations employed the PSpice model of the operational amplifier, providing accurate numerical predictions. The experimental results were validated against these simulations. All components, including capacitors and inductors, had tolerances ranging from 1% to 5%, ensuring the reliability of the circuit's performance. The high precision in component values and the careful design of additional elements guaranteed the stability and accuracy of the experimental setup.

## 3. Results

*3.1 Non-Hermitian reshaping of high-order Landau modes in continuum model*

To illustrate the mechanism of non-Hermitian reshaping of Landau modes, we begin with a one-dimensional (1D) non-Hermitian harmonic oscillator model, which is given by

$$H = q_y^2 + y^2 + 2i\gamma q_y \tag{1}$$

where $y$ denotes the coordinate operator, $q_y$ is the momentum operator, and the term $2i\gamma q_y$ introduces an imaginary momentum. The eigen energies can be analytically derived as $\omega_n = 2n + \gamma^2$, where $n = 0,1,2 \cdots$ is the energy index. The corresponding eigenstate is $\phi_n = CH_n(y)e^{-y^2/2}e^{\gamma y}$, where $H_n$ is the $n$-th order Hermite polynomial and $C$ is the normalization constant (The detailed derivation can be found in Note S1). For the Hermitian case with $\gamma = 0$, the spatial profile of harmonic oscillator modes has two features: (i) it exhibits $n + 1$ peaks due to the $n$ zeros of the $n$-th order Hermite polynomial; and (ii) it is symmetric about $y = 0$ due to the even parity of $|\phi_n|^2$, and its distribution diffuses with increasing quantum number $n$, as shown in the left panel of Fig. 1a. Interestingly, the symmetry of the harmonic oscillator modes is broken when $\gamma \neq 0$. This leads to a gradual evolution from multiple peaks into a single peak profile. Furthermore, the localization center at $y = 0$ is displaced along the $y$-direction as $\gamma$ increases. Such a Hamiltonian can be realized in the Hatano-Nelson model [24] subjected to a quadratic potential modulation. (See more details in Note S2.)

The above 1D non-Hermitian harmonic oscillator model provides insight into how imaginary momentum reshapes the harmonic oscillator modes, while more intriguing phenomena emerge in two-dimensional (2D) systems. The extension to 2D systems gives rise to Landau modes, whose high degeneracy and topological properties render them an ideal platform for mode reshaping. The 2D system is also particularly fertile for exploring rich non-Hermitian physics [20,21,25–37]. Here, we consider a continuum model for a 2D Dirac system (here we set $\hbar = e = 1$) subjected to an electric field $E_x$, a magnetic field $B_z$, and an imaginary momentum $i\gamma$ described by the Hamiltonian

$$H = E_x x \sigma_0 + v_x(q_x - B_z y)\sigma_x + v_y(q_y + i\gamma)\sigma_y \tag{2}$$

where $v_x$ and $v_y$ are Fermi velocities, and $\sigma_i (i = x, y)$ denote Pauli matrices and $\sigma_0$ is the identity matrix.

First, we consider the purely magnetic field case ($B_z \neq 0$, $E_x = 0$, and $\gamma = 0$). The continuous spectrum is quantized into discrete Landau levels with energies $\omega_n^{(0)} = \text{sgn}(n)\sqrt{|2v_x v_y n B_z|}$, where $n = 0, \pm 1, \pm 2 \cdots$, as shown in Fig. 1b and c. The corresponding Landau modes are given by $\psi_{n,q_x}^{(0)} \propto e^{iq_x x} \begin{pmatrix} \phi_{|n|}\left(\frac{1}{l_c'}\frac{\sqrt{v_x}}{\sqrt{v_y}}(y - y_c)\right) \\ \text{sgn}(n)\phi_{|n-1|}\left(\frac{1}{l_c'}\frac{\sqrt{v_x}}{\sqrt{v_y}}(y - y_c)\right) \end{pmatrix}$, $\phi_{|n|}$ and $\phi_{|n-1|}$ is the two-component spinor. Here, $\phi_{|n|}$ is the $n$-th order harmonic oscillator wavefunction, $y_c = q_x/B_z$ and $l_c' = \sqrt{1/|B_z|}$ is the magnetic length. The detailed derivation can be found in Note S3. The Landau mode is a product of a 1D harmonic oscillator state in the $y$-direction and a plane wave $e^{iq_x x}$ in the $x$-direction, as shown in Fig. 1d(I) and e(I). In a system of finite length $L_x$, by imposing periodic boundary conditions $\psi_{|n|,q_x}^{(0)}(0) = \psi_{|n|,q_x}^{(0)}(L_x)$, the momentum is quantized as $q_{x_j} = 2\pi j/L_x$ ($j = 1, 2, \cdots s$), where $s$ is the degeneracy of each Landau level. Since all these $s$ states share the same energy, this leads to $s$-fold degeneracy for each Landau level. This degeneracy provides a promising platform for information processing.

Next, we analyze the influence of the imaginary momentum $\gamma$ on the Landau modes ($B_z \neq 0, E = 0$, and $\gamma \neq 0$). The imaginary momentum modifies the Landau modes to $e^{\gamma y}\psi_{|n|}^{(0)}$ (see more details in Note S3). This breaks the reflection symmetry previously centered at $y = y_c$. Consequently, the Landau modes evolve from multiple peaks into a single peak and their localization centers are separated, as shown in Fig. 1 d(II) and e(II).

Furthermore, we investigate the effect of an electric field ($B_z \neq 0$, $E_x \neq 0$, and $\gamma = 0$). The electric potential $V = E_x x$ acts as a perturbation and lifts the $s$-fold degeneracy of the Landau levels as shown in Fig. 1f. The new Landau modes, denoted as $\psi_{n,m}$, are linear combinations of the original degenerate Landau modes $\psi_{n,q_{x_j}}^{(0)}$.

$$\psi_{n,m} = \sum_j^s c_{m,j} \psi_{n,q_{x_j}}^{(0)}\left(y - q_{x_j}/B_z\right) \tag{3}$$

The expansion coefficients $c_{m,j}$ are derived by diagonalizing the perturbation Hamiltonian $V$ within the degenerate subspace of the $n$-th Landau level. We apply first-order degenerate perturbation theory to obtain the new spectrum. This approach is valid as long as the energy correction from the electric field is much smaller than the gap between adjacent Landau levels. Crucially, our system works within the "magnetic regime" where the effective drift velocity $E_x/B_z$ is smaller than the Fermi velocity [38]. Distinct from Wannier-Stark localization, which dominates in the strong-electric-field regime [39], the electric field in our system acts as a weak perturbation that lifts the guiding-center degeneracy of each Landau level. The spectrum is given by (see more details in Note S4)

$$\omega_{n,m} = \omega_n^{(0)} + E_x \langle n,m|x|n,m \rangle, m = 1,2,\cdots,s \tag{4}$$

Remarkably, the degeneracy of the Landau modes is broken, and energy splitting is directly proportional to the position expectation value $\langle x \rangle_{n,m}$. Thus, the Landau modes with different energies are now located at different positions in the $x$-direction, as shown in Fig. 1d(III) and e(III)

We note that the energy-position correspondence can also be derived in the alternative gauge $\mathbf{A} = (0, B_z x, 0)$ via a Lorentz boost [39]. However, in the presence of a pseudo-magnetic field, the Landau modes in time-reversal symmetry related valleys have opposite group velocities ($v_g^{\pm} = \mp E_x/B_z$), while sharing the same energy and guiding center. The inter-valley scattering forms a standing-wave pattern. Therefore, the gauge $\mathbf{A} = (-B_z y, 0, 0)$ is more suitable for realizing the guiding-center splitting mechanism in the presence of a pseudo-magnetic field.

Finally, we consider the case where all three components are present ($B_z \neq 0$, $E_x \neq 0$, and $\gamma \neq 0$). The interplay of these components reveals rich phenomena. The degeneracy of the Landau modes is lifted in energy, and the modes are spatially separated along the *x*-direction under the influence of electric field. Furthermore, each mode now exhibits a single peak profile along the *y*-direction due to the imaginary momentum. As a concrete example of non-Hermitian reshaping of high-order Landau modes, we present a non-Hermitian Landau mode reconfiguration phenomenon, as shown in Fig. 1d(IV) and e(IV).

*3.2 Non-Hermitian reshaping of high-order Landau modes in Lattice model*

To verify the above mechanism in realistic systems, we consider a honeycomb lattice, as shown in Fig. 2a. We construct a PMF by engineering an inhomogeneous coupling along the *y*-direction, and generate an imaginary momentum $\gamma$ through non-reciprocal coupling. The tight-binding Hamiltonian is

$$H = \sum_r \sum_{n=1}^{3} \left( t_n \hat{a}_r^\dagger \hat{b}_{r-\vec{\delta}_n} + t_n' \hat{a}_r \hat{b}_{r-\vec{\delta}_n}^\dagger \right) \quad (5)$$

where $\vec{\delta}_{1,2,3}$ are the nearest-neighbor vectors, with $\vec{\delta}_1 = (0,1)$, $\vec{\delta}_2 = (-\sqrt{3}/2, -1/2)$ and $\vec{\delta}_3 = (\sqrt{3}/2, -1/2)$; $\hat{a}_r^\dagger(\hat{a}_r)$ and $\hat{b}_r^\dagger(\hat{b})$ are the creation (annihilation) operators and $t_{1,2,3}$ ($t_{1,2,3}'$) are coupling coefficients along the $\vec{\delta}_{1,2,3}$ directions, respectively. Here, the nearest-neighbor spacing is set to unity. We set $t_{1,2} = t'_{1,2} = t$ to be uniform, while the coupling along the $\vec{\delta}_3$ direction is modulated by $t_3 = t(1 + \alpha y) + \gamma$ and $t'_3 = t(1 + \alpha y) - \gamma$ for a ribbon chain with $\alpha = 1.5/N_y$, where $N_y$ is the number of unit cells in the $y$-direction. Near the $K(K')$ valleys, our lattice model is consistent with the above continuum model of Eq. (2) (see more details in Note S5). The valley-dependent artificial gauge potential is $\vec{A} = (\xi 2\alpha y/3, 0)$, where $\xi = \pm 1$ distinguishes the $K/K'$ valleys; the PMF is given by $B_z = \nabla \times \vec{A} = \xi 2\alpha/3$; and the Fermi velocities are $v_x = 3t/2$ and $v_y = 3t/2(1 + 2\alpha y/3)$. As a direct consequence of this gauge potential, the Dirac points in different valleys shift along opposite directions as $N_y$ increases, as shown in Fig. 2b. Furthermore, the PMF splits the spectrum into discrete Landau levels with energies $\omega_n = \text{sgn}(n)t\sqrt{3|n|\alpha(1 + \xi q_x)}$. In addition, due to the non-uniform Fermi velocity $v_y = 3at/2(1 + 2\alpha y/3)$ along the $y$-direction, the higher-order ($n>0$) Landau levels are not flat.

To quantify the localization properties of Landau modes, we calculate the participation ratio (PR) [40] $PR = (\sum_r |\psi_r|^2)^2 / \sum_r |\psi_r|^4$, where $|\psi\rangle$ is the eigenstate for a given frequency and $|\psi_r| = |\langle r|\psi\rangle|$ is the amplitude at site $r$. A large value of the PR corresponds to a delocalized state, whereas a small value indicates strong localization. As the Landau level index $|n|$ increases, the corresponding PR becomes larger, indicating that the high-order Landau modes are more spatially extended, as shown in Fig. 2c.

Then, we introduce an imaginary momentum ($\gamma \neq 0$) via the non-reciprocal coupling [41–44]. Compared to the purely Hermitian case ($\gamma = 0$), the introduction of imaginary momentum ($\gamma = 0.2$) leads to a significant decrease in the PR for high-order Landau modes. Thus, the sharp drop in PR indicates that the high-order Landau modes become strongly localized in space, as shown in Fig. 2d. It is worth noting that the non-Hermitian skin effect is suppressed by the pseudo-magnetic field (see Note S6). Thus, the reduction of PR originates from the reshaping of high-order Landau modes rather than the non-Hermitian skin effect. However, the high-order Landau modes have the same frequency but opposite velocities between $K/K'$ valleys. These degenerate states are susceptible to scattering between different valleys, which hinders their practical application. In order to decrease the scattering [45–47] and reshape high-order Landau modes in finite structures, we simultaneously construct a PEF ($E_x$) by spatially modulating the onsite potential along the $x$–direction $V(x) = E_x x$ (Fig. 2e), and introduce an non-reciprocal coupling ($\gamma$). For the numerical calculations on the finite structure ($N_y = 30, N_x = 20$), the parameters are chosen as $\gamma = 0.3$, $\alpha = 1.8/N_y$, and $E_x = 0.12/x$, respectively. Thus, the high-order Landau modes with different frequencies are localized at different $x$ positions, as shown in Fig. 2f.

To visualize the distribution of high-order Landau modes, we plot the squared modulus of the different wavefunctions, as shown in Fig. 3a–f. To improve clarity, we show the profiles of alternate Landau modes in Fig. 3 to avoid overlap. The full set of Landau modes is provided in Fig. S8. When only PMF is present, the Landau modes are extended along the $x$-direction due to the valley degeneracy, as shown in Fig. 3a and b. When PEF is also applied, the valley degeneracy of high–order Landau modes is lifted. Thus, the modes separate along the $x$–direction but exhibit $n + 1$ peaks in the $y$-direction, as shown in Fig. 3c and d. Finally, the combined action of PMF, PEF, and $\gamma$ reshapes the high-order Landau modes by transforming them into a single peak along the $y$-direction while simultaneously separating them into different $x$-positions according to their frequency, as shown in Fig. 3e and f. The individual modes shown in Fig. 3e and f correspond to the eigenstates from $n = 1$ and $n = 2$ Landau levels of Fig. 2f.

To investigate the ability of PEF and $\gamma$ to reshape the high-order Landau modes in a finite lattice model, we select the $n = 2$ Landau modes at fixed frequency $\omega = 0.68$ for our study. First, we consider the influence of PEF on these mode by setting $\gamma = 0$. As PEF increases, the spatial extent of the mode along the $x$-direction decreases (Fig. 3g), which indicates that PEF breaks valley degeneracy and suppresses inter-valley scattering. Next, the PEF is set to be $E_x = 0.004$, and we examine the effect of different values of $\gamma$. As $\gamma$ increases, the mode gradually transforms into a single peak in the *y*-direction and shifts towards the upper boundary, as shown in Fig. 3h. To test the robustness of the reshaped Landau modes, we introduce disorder ($\Delta$) into the coupling $t_3 = t(1 + \alpha(1 + \Delta)y)$. We find that the localization properties of the reshaped Landau modes are robust even when the disorder exceeds the typical experimental error range (i.e., 5%). Within this range, the non-Hermitian Landau modes reconfiguration phenomenon also remains robust. The robust properties are advantageous for application in frequency multiplexing and wave packet reshaping (The detailed derivation can be found in Note S7).

*3.3 Observation of the non-Hermitian reshaping of high-order Landau modes in electric circuits*

As for experimental verification, we built an electric circuit platform based on the above designed 2D honeycomb lattice model, as illustrated in Fig. 4a. The lattice sites are labeled as $a_{m,n}$ and $b_{m,n}$, where $m$ and $n$ denote the unit cell indices in the *x*– and *y*–directions, respectively. The intracell coupling strength is represented by the capacitor $C_0$, corresponding to the hopping amplitude $t$. The intercell coupling along the *y*-direction is denoted as $C_n$, which increase linearly with $n$: $C_n = C_0 + nC_1$, thus implementing $t + \alpha y$. The change in coupling generates an effective gauge potential. Meanwhile, a PEF is implemented by adjusting the on-site potential along the *x*-direction (i.e., by tuning the capacitances $C_x$ of the corresponding nodes). Imaginary momentum hopping is achieved by using voltage followers (VFs) with an additional capacitance $\delta C$ on each link, corresponding to $2\gamma$ in the Hamiltonian. At the fixed frequency $\omega$, the circuit Laplacian can be written as (which can be seen in Note S8)

$$J(\omega)/(-i\omega) = \sum_{\vec{r}}(t_1 + \alpha r_y \hat{y} + 2\gamma)a_{\vec{r}}^\dagger b_{\vec{r}-\vec{\delta}_1} + (t_1 + \alpha r_y \hat{y})\sum_{\vec{r}} a_{\vec{r}} b_{\vec{r}-\vec{\delta}_1}^\dagger + t_2 \sum_{\vec{r}} \left(a_{\vec{r}}^\dagger b_{\vec{r}-\vec{\delta}_2} + \right.$$

$$\left. a_{\vec{r}} b_{\vec{r}-\vec{\delta}_2}^\dagger \right) + t_3 \sum_{\vec{r}} \left(a_{\vec{r}}^\dagger b_{\vec{r}-\vec{\delta}_3} + a_{\vec{r}} b_{\vec{r}-\vec{\delta}_3}^\dagger \right) + \sum_{\vec{r}} (\varepsilon_0 - E r_x \hat{x})(a_{\vec{r}}^\dagger a_{\vec{r}} + b_{\vec{r}}^\dagger b_{\vec{r}}) \tag{6}$$

where $t_1 = t_2 = t_3 = C_0$, $2\gamma = \delta C$, $\alpha = C_1$, $\varepsilon_0 = 1/(\omega^2 L_0) - C_{x0} - 3C_0$ and $E_x = \beta C_0$, with $\beta$ being the modulation factor of the on-site potential that characterizes the strength of the PEF $E_x = \beta C_0$. Figure. 4b presents the experimental results of the real admittance spectrum together with the PR at the resonant frequency $f$ = 205.5 kHz. The error bars represent the standard deviation obtained from multiple experimental measurements. The experimental data agree well with the theoretical predictions, clearly demonstrating the expected spatial and frequency separation. To gain deeper insight into the underlying mechanisms, we simulate high-order Landau modes under different PEF and imaginary momentum to elucidate how these factors reshape them. Figure. 4c illustrates that the PEF reduces the overlap of counter-propagating Landau modes, which become better separated in the $x$-direction with increasing $\beta$. On the other hand, Fig. 4d shows the effect of increasing the non-Hermitian parameter $\delta C$ (from 0 to 16 nF). As $\delta C$ increases, the peaks of Landau modes gradually decrease in the $y$-direction, reflecting how non-Hermitian modulation alters the envelope and energy concentration of Landau modes. The details can be found in Note S9.

As a concrete example of non-Hermitian reshaping of high-order Landau modes, we directly observed a non-Hermitian Landau mode reconfiguration and compared the results with theoretical simulations. For simplicity, we chose $m = 5$ and $n = 14$ for the structure design in our experiment, which was fabricated on a $30 \times 50$ cm$^2$ circuit board. To experimentally verify the spatial profiles of the Landau modes, we excited the physical circuit by applying AC signals to the nodes exhibiting maximum modal amplitudes at their resonance frequencies, and subsequently measured the voltage response across all nodes. Theoretical calculations (Fig. 4e) predict that for the first-order Landau level ($n = 1$), the Landau modes should be delocalized due to valley degeneracy. This was confirmed in our experiment when only the PMF was present, as shown by the delocalized modes in Fig. 4f. In contrast, when PMF, PEF, and $\gamma$ are simultaneously introduced, the theoretical model predicts that the Landau modes transform into highly localized states separated in both space and frequency (Fig. 4g). This reconfiguration was successfully observed in our experiment, where the excited modes became highly localized (Fig. 4h).

Similarly, for the second-order Landau level ($n = 2$), when PMF, PEF, and $\gamma$ were simultaneously introduced, the Landau modes also separated in both space and frequency. In contrast to the delocalized PMF case (Fig. 4i and j), these modes exhibited single peak profiles (Fig. 4k and l) Collectively, these figures illustrate the voltage distributions of the Landau modes under the combined effects of PMF, PEF, and imaginary momentum coupling at different frequencies and excitation positions, clearly revealing the spatial localization and frequency dispersion characteristics of the Landau mode reconfiguration realized in our experimental circuit platform.

## Conclusion

In conclusion, we successfully reshape the high-order Landau modes in non-Hermitian systems both in theory and experiment, addressing the key challenges of their spatial diffuse and degenerate properties. Our work provides a universal analytical framework based on the joint action of magnetic fields, electric fields, and an imaginary momentum to realize the frequency-dependent localization of high-order Landau modes. It establishes a direct mapping between the frequency and the spatial profile of the reshaped high-order Landau modes. Experimentally, we realize these effects by building an electric circuit platform that implements a PMF, a PEF, and non-reciprocal coupling, and directly observe the frequency-dependent spatial localization phenomenon of the high-order Landau modes on this platform. The underlying physical mechanism of this phenomenon is universal and can be extended to other systems, such as photonic, acoustic, and elastic systems, demonstrating that the combination of artificial gauge fields and non-Hermiticity enables versatile and effective control of Landau modes to meet diverse functional requirements. Our work not only presents a novel idea to manipulate Landau modes but also opens avenues for exploring potential applications such as frequency multiplexing and wave packet reshaping.

## Conflict of interest

The authors declare that they have no conflict of interest.

## Acknowledgments

This work was supported by the National Key R&D Program of China (2023YFA1407600), the National Natural Science Foundation of China (12574334, 12274031, 92576204), the Fundamental Research Funds for the Central Universities (2025CX01017), the New Cornerstone Science Foundation, the Research Grants Council of Hong Kong (STG3/E-704/23-N, AoE/P-502/20 and 17309021), Guangdong Provincial Quantum Science Strategic Initiative (GDZX2204004, GDZX2304001).

## Author contributions

Zhihao Wang performed theoretical calculations and numerical simulations. Jie Jiang performed simulations of the electric circuit. Zhiwei Guo and JieJiang performed experiments. Yanji Zheng, Wen Zhao, and Chenyang Wang participated in the theoretical calculations and numerical simulations. Cuicui Lu, Yong-Chun Liu, and Shuang Zhang analyzed the results. Cuicui Lu, Zhihao Wang, Jie Jiang, Yong-Chun Liu and Shuang Zhang wrote the manuscript. Cuicui Lu, Shuang Zhang, Yong-Chun Liu and Zhiwei Guo supervised the projects. All the authors contributed to discuss and review the manuscript.

## Appendix A. Supplementary materials

Supplementary materials to this article can be found online at

## References

[1]  Klitzing KV, Dorda G, Pepper M. New method for high-accuracy determination of the fine-structure constant based on quantized Hall resistance. Phys Rev Lett 1980;45:494–497.

[2] Gusynin VP, Sharapov SG. Unconventional integer quantum Hall effect in graphene. Phys Rev Lett 2005;95:146801.

[3] Zhang Y, Tan YW, Stormer HL, et al. Experimental observation of the quantum Hall effect and Berry's phase in graphene. Nature 2005;438:201–204.

[4] Li G, Andrei EY. Observation of Landau levels of Dirac fermions in graphite. Nat Phys 2007;3:623–627.

[5] Guinea F, Katsnelson MI, Geim AK. Energy gaps and a zero-field quantum Hall effect in graphene by strain engineering. Nat Phys 2010;6:30–33.

[6] Levy N, Burke SA, Meaker KL, et al. Strain-induced pseudo–magnetic fields greater than 300 Tesla in graphene nanobubbles. Science 2010;329:544–547.

[7] Pereira VM, Castro Neto AH, Peres NMR. Tight-binding approach to uniaxial strain in graphene. Phys Rev B 2009;80:045401.

[8] Rechtsman MC, Zeuner JM, Tünnermann A, et al. Strain-induced pseudomagnetic field and photonic Landau levels in dielectric structures. Nat Photonics 2013;7:153–158.

[9] Yang Z, Gao F, Yang Y, et al. Strain-induced gauge field and Landau levels in acoustic structures. Phys Rev Lett 2017;118:194301.

[10] Wen X, Qiu C, Qi Y, et al. Acoustic Landau quantization and quantum-Hall-like edge states. Nat Phys 2019;15:352–356.

[11] Barsukova M, Grisé F, Zhang Z, et al. Direct observation of Landau levels in silicon photonic crystals. Nat Photonics 2024;18:580–585.

[12] Barczyk R, Kuipers L, Verhagen E. Observation of Landau levels and chiral edge states in photonic crystals through pseudomagnetic fields induced by synthetic strain. Nat Photonics 2024;18:574–579.

[13] Zhao W, Zheng Y, Lu C, et al. Landau rainbow induced by artificial gauge fields. Phys Rev Lett 2024;133:233801.

[14] Brendel C, Peano V, Painter OJ, et al. Pseudomagnetic fields for sound at the nanoscale. Proc Natl Acad Sci USA 2017;114:E3390–E3395.

[15] Yan M, Deng W, Huang X, et al. Pseudomagnetic fields enabled manipulation of on-chip elastic waves. Phys Rev Lett 2021;127:136401.

[16] Wang W, Gao W, Chen X, et al. Moiré fringe induced gauge field in photonics. Phys Rev Lett 2020;125:203901.

[17] Cheng Z, Guan YJ, Xue H, et al. Three-dimensional flat Landau levels in an inhomogeneous acoustic crystal. Nat Commun 2024;15:2174.

[18] Mann CR, Horsley SAR, Mariani E. Tunable pseudo-magnetic fields for polaritons in strained metasurfaces. Nat Photonics 2020;14:669–674.

[19] Yang J, Li Y, Yang Y, et al. Realization of all-band-flat photonic lattices. Nat Commun 2024;15:1484.

[20] Wang Q, Zhu C, Zheng X, et al. Continuum of bound states in a non-Hermitian model. Phys Rev Lett 2023;130:103602.

[21] Zhang X, Wu C, Yan M, et al. Observation of continuum Landau modes in non-Hermitian electric circuits. Nat Commun 2024;15:1798.

[22] Jia H, Zhang R, Gao W, et al. Observation of chiral zero mode in inhomogeneous three-dimensional Weyl metamaterials. Science 2019;363:148–151.

[23] Teo HT, Mandal S, Long Y, et al. Pseudomagnetic suppression of non-Hermitian skin effect. Sci Bull 2024;69:1667–1673.

[24] Hatano N, Nelson DR. Localization transitions in non-Hermitian quantum mechanics. Phys Rev Lett 1996;77:570–573.

[25] Shao K, Cai ZT, Geng H, et al. Cyclotron quantization and mirror-time transition on nonreciprocal lattices. Phys Rev B 2022;106:L081402.
[26] Zhang X, Tian Y, Jiang JH, et al. Observation of higher-order non-Hermitian skin effect. Nat Commun 2021;12:5377.
[27] Zou D, Chen T, He W, et al. Observation of hybrid higher-order skin-topological effect in non-Hermitian topolectrical circuits. Nat Commun 2021;12:7201.
[28] Li CA, Trauzettel B, Neupert T, et al. Enhancement of second-order non-Hermitian skin effect by magnetic fields. Phys Rev Lett 2023;131:116601.
[29] Zhao E, Wang Z, He C, et al. Two-dimensional non-Hermitian skin effect in an ultracold Fermi gas. Nature 2025;637:565–573.
[30] Lee CH, Li L, Gong J. Hybrid higher-order skin-topological modes in nonreciprocal systems. Phys Rev Lett 2019;123:016805.
[31] Sun Y, Hou X, Wan T, et al. Photonic Floquet skin-topological effect. Phys Rev Lett 2024;132:063804.
[32] Li Y, Liang C, Wang C, et al. Gain-loss-induced hybrid skin-topological effect. Phys Rev Lett 2022;128:223903.
[33] Xie X, Ma F, Rui WB, et al. Non-Hermitian Dirac cones with valley-dependent lifetimes. Nat Commun 2025;16:1627.
[34] Xue H, Wang Q, Zhang B, et al. Non-Hermitian Dirac cones. Phys Rev Lett 2020;124:236403.
[35] Zhang K, Yang Z, Fang C. Universal non-Hermitian skin effect in two and higher dimensions. Nat Commun 2022;13:2496.
[36] Hu H. Topological origin of non-Hermitian skin effect in higher dimensions and uniform spectra. Sci Bull 2025;70:51–57.
[37] Wan T, Zhang K, Li J, et al. Observation of the geometry-dependent skin effect and dynamical degeneracy splitting. Sci Bull 2023;68:2330–2335.
[38] Lukose V, Shankar R, Baskaran G. Novel electric field effects on Landau levels in graphene. Phys Rev Lett 2007;98:116802.
[39] Wannier GH. Wave functions and effective Hamiltonian for Bloch electrons in an electric field. Phys Rev 1960;117:432–439.
[40] Thouless DJ. Electrons in disordered systems and the theory of localization. Phys Rep 1974;13:93.
[41] Brandenbourger M, Locsin X, Lerner E, et al. Non-reciprocal robotic metamaterials. Nat Commun 2019;10:4608.
[42] Helbig T, Hofmann T, Imhof S, et al. Generalized bulk–boundary correspondence in non-Hermitian topolectrical circuits. Nat Phys 2020;16:747–750.
[43] Weidemann S, Kremer M, Helbig T, et al. Topological funneling of light. Science 2020;368:311–314.
[44] Wang W, Wang X, Ma G. Non-Hermitian morphing of topological modes. Nature 2022;608:50–55.
[45] Salerno G, Ozawa T, Price HM, et al. How to directly observe Landau levels in driven-dissipative strained honeycomb lattices. 2D Mater 2015;2:034015.
[46] Lantagne-Hurtubise É, Zhang XX, Franz M. Dispersive Landau levels and valley currents in strained graphene nanoribbons. Phys Rev B 2020;101:085423.
[47] Bao Z, Ding J, Qi J. Complex Landau levels and related transport properties in the strained zigzag graphene nanoribbons. Phys Rev B 2023;107:125411.

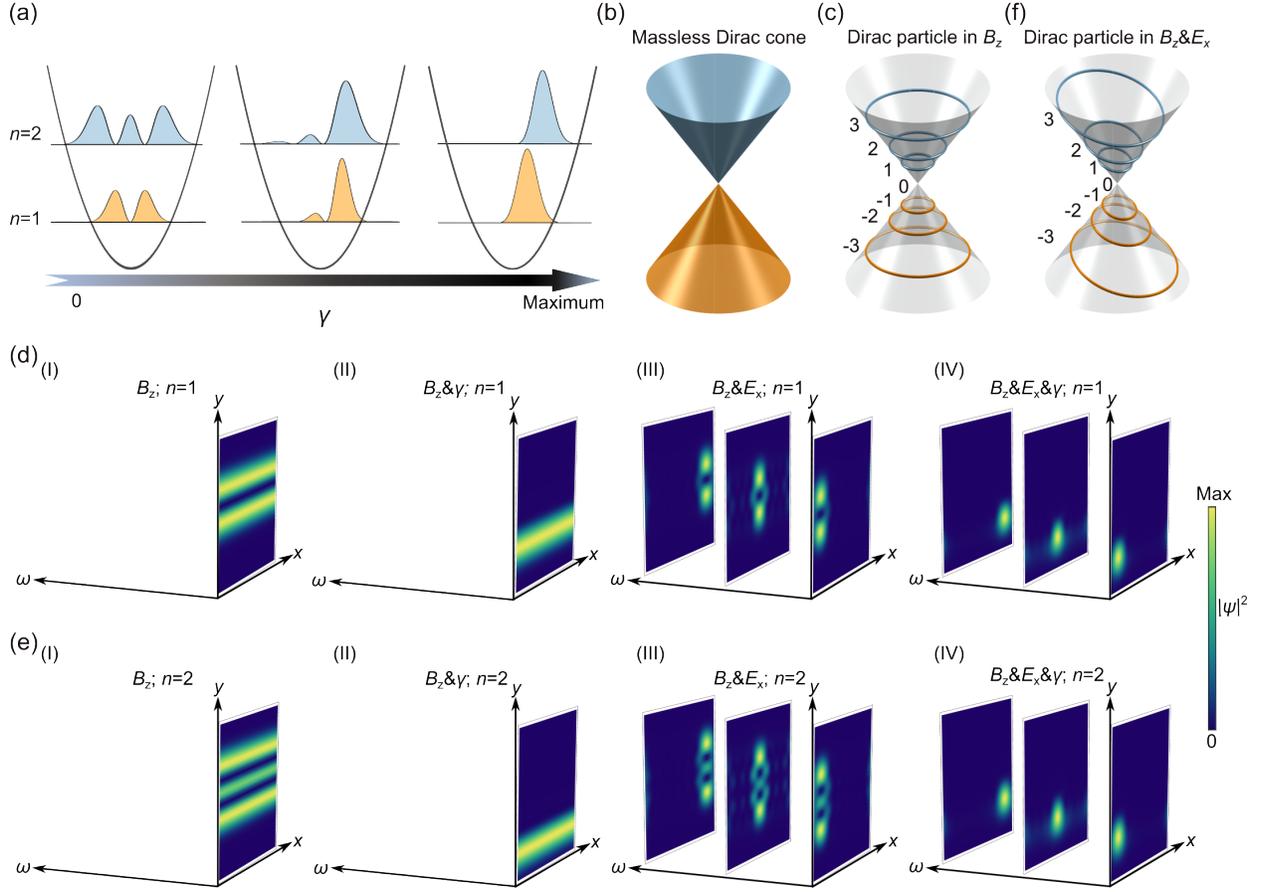

**Fig. 1** Schematic of non-Hermitian reshaping of Landau modes. (a) Schematic of non-Hermitian reshaping of $n = 1$ and $n = 2$ Landau modes in a one-dimensional case. The black arrows indicate increasing imaginary momentum. (b, c) Schematic of a massless Dirac cone splitting into Landau levels and the degeneracy breaking of Landau levels in a two-dimensional case. (b) The spectrum for the standard Dirac system. (c) Formation of a series of Landau levels in the Dirac system under a magnetic field $B_z$. (d–e) Spatial profiles of $n = 1$ and $n = 2$ Landau modes under different conditions, calculated using the Landau gauge $(B_z y, 0, 0)$. (d)(I) and (e)(I) Landau modes are degenerate in frequency when only $B_z$ exists. The spatial distributions of these Landau modes are described by 1D Hermitian harmonic oscillator modes along the $y$-direction and plane waves along the $x$-direction. (d)(II) and (e)(II) When both $B_z$ and $\gamma$ are present, the Landau modes remain degenerate in energy. However, their spatial profiles along the $y$-direction transform from multi-peak to single-peak structures, while maintaining plane waves in the $x$-direction. (d)(III) and (e)(III) When $B_z$ and $E_x$ are present, the degeneracy is lifted, and the Landau modes with different energies become localized at different positions along the $x$-direction. (d)(IV) and (e)(IV) When $B_z$, $E_x$, and $\gamma$ are all present, the degeneracy is lifted. The Landau modes form single-peak profiles along the

*y*-direction and separate along the *x*-direction. (f) The degeneracy of the Landau levels is lifted under a magnetic field $B_z$ and an electric field $E_x$.

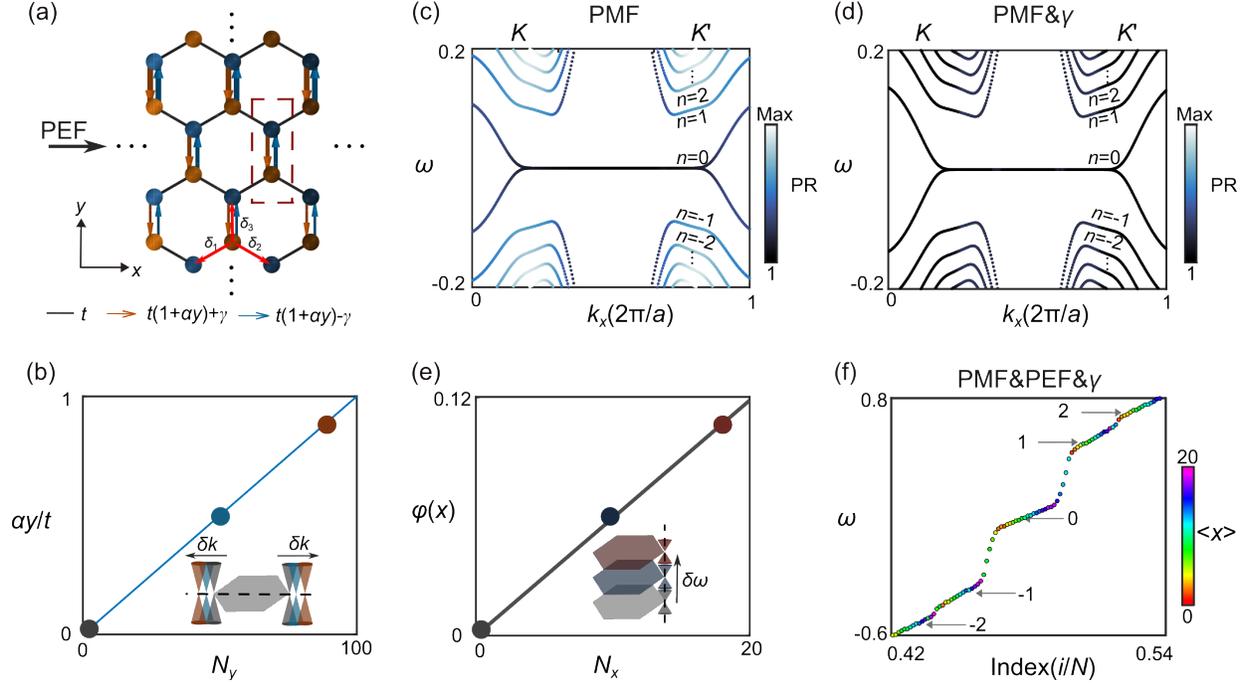

**Fig. 2** Non-Hermitian reshaping of high-order Landau modes in a lattice model. (a) Schematic of a honeycomb lattice model. Coupling strengths in the $y$-direction are given by $t+\alpha y\pm\gamma$. The atomic color, ranging from light to dark, represents increasing on-site potential in $x$-direction. (b) The coupling variation of the supercell along the $y$-direction. The insets show the trajectory of the Dirac point as the $y$ position increases. (c) Energy spectrum with periodic boundary conditions along $x$-direction and open boundary conditions along the $y$-direction when only a PMF exists. The system consists of $N_y = 100$ unit cell along the $y$-direction with periodic boundary conditions in the $x$-direction. The color represents the value of PR. (d) Energy spectrum with periodic boundary conditions along $x$-direction and open boundary conditions along the $y$-direction when PMF and $\gamma = 0.2$ both exist. (e) Variation of the on-site potential along $x$-direction. The inset illustrates that the Dirac point frequency increases as the on-site potential grows along $x$-direction. (f) Energy spectrum under open boundary conditions along both the $x$- and $y$-directions. The system consists of $N_y = 30$ unit cells along the $y$-direction and $N_x = 20$ unit cells along $x$-direction. We set $\alpha = 1.8/N_y$ and $\gamma = 0.2$ for the calculation. The rainbow colors indicate the position expectation value $<x>$ of each eigenstate.

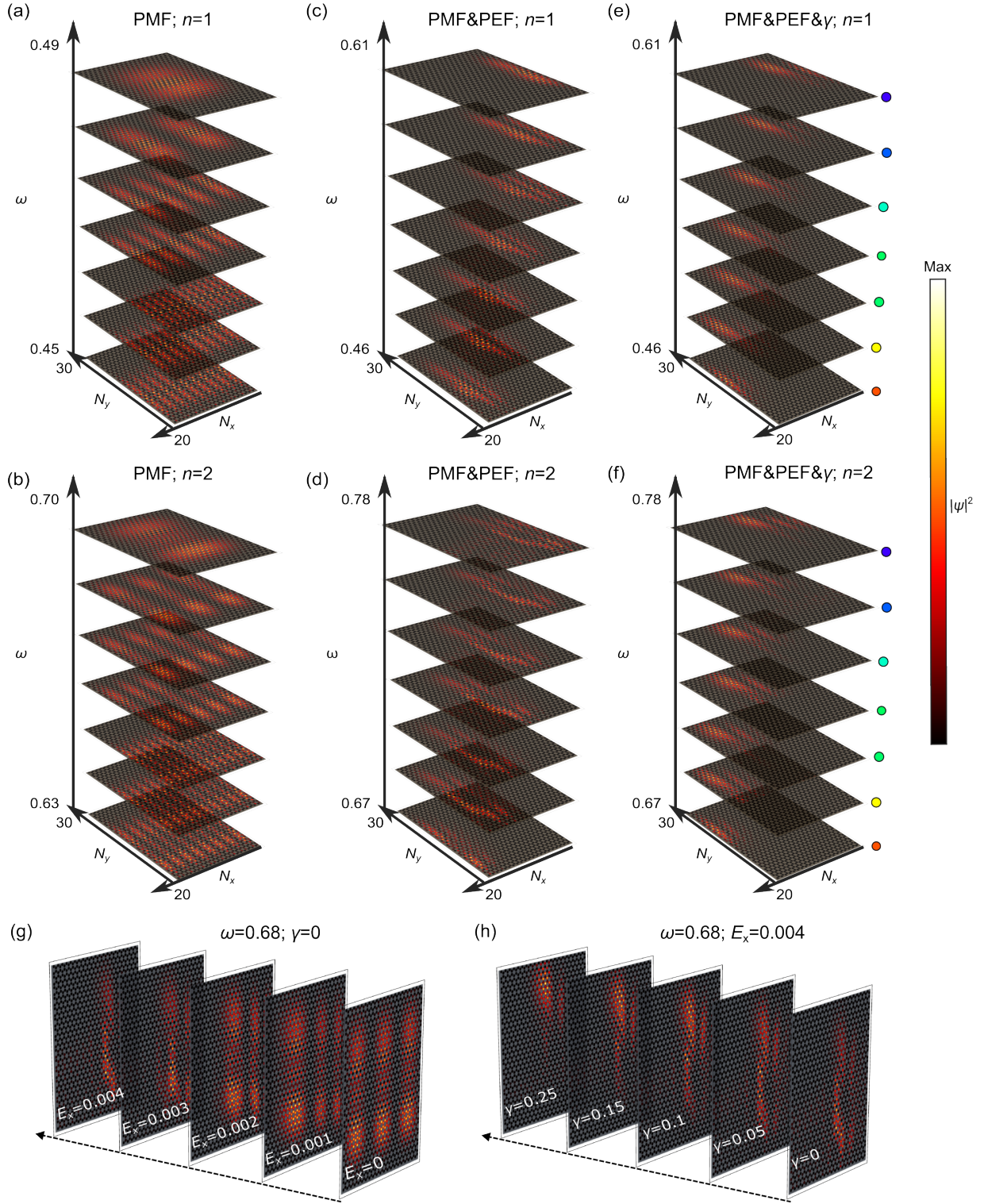

**Fig. 3** Reshaping of Landau modes in a non-Hermitian lattice model. Landau modes corresponding to the $n = 1$ and $n = 2$ Landau levels under three kinds of conditions: (a, b) When only PMF is present, the Landau modes exhibit spatial overlap. (c, d) The PEF lifts the valley degeneracy of the Landau modes. The Landau modes are separated both in frequency $\omega$ and in space along the $x$-direction. (e, f) With a PMF, a PEF, and non-reciprocal coupling $\gamma$ all present, the Landau modes separated both in frequency ω and in space along the $x$-direction, and the mode profiles are reshaped into highly localized single peaks in the $y$-direction. The colormap is consistent with the energy spectrum shown in Fig. 2f. (g) The Landau modes under different values of PEF. (h) The Landau modes under different values of $\gamma$.

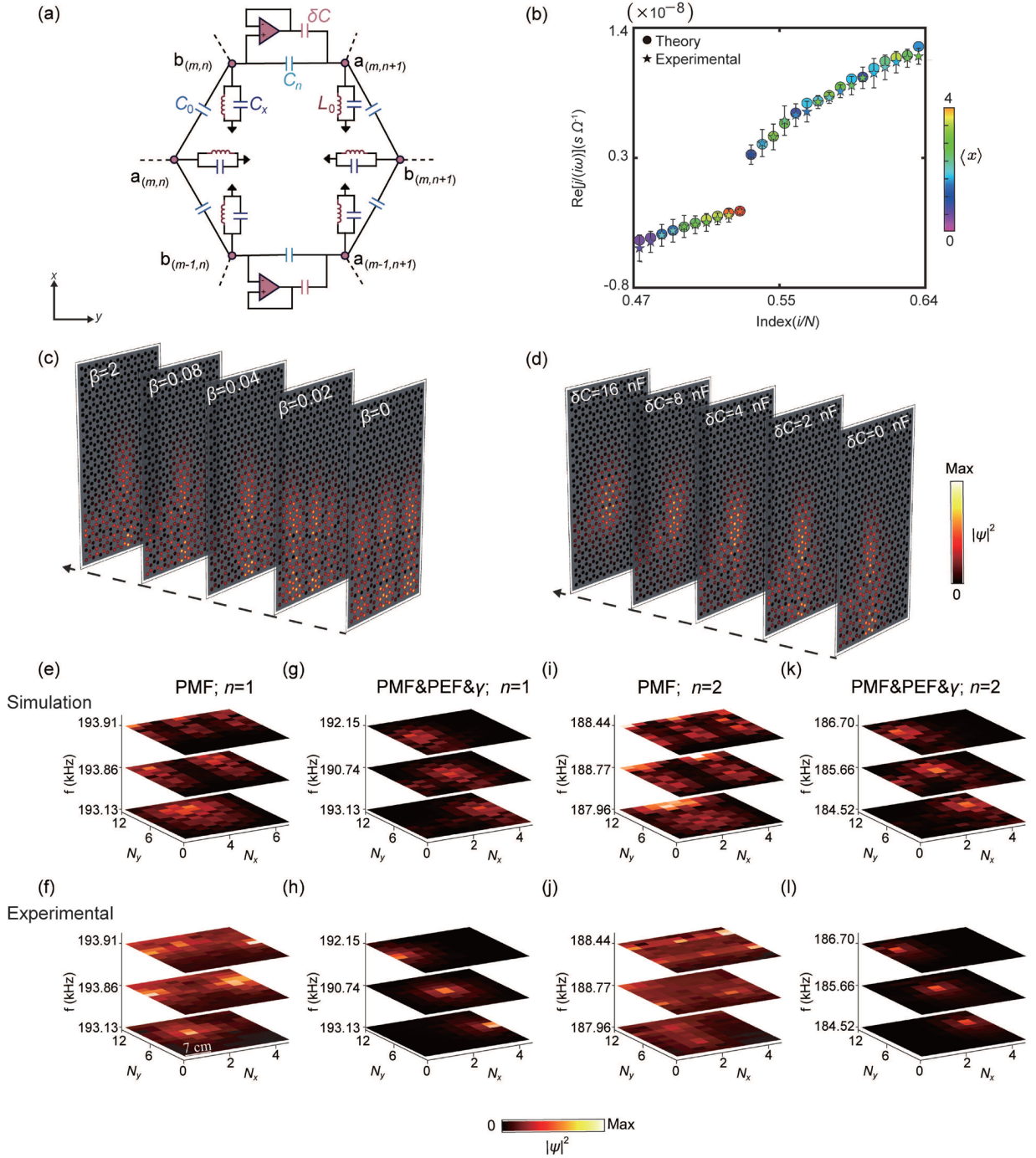

**Fig. 4** Experimental observation of the reshaping of Landau modes in non-Hermitian electric circuit systems. (a) Schematic of one unit of circuit honeycomb lattices, labeled by a$_{m,n}$ and b$_{m,n}$. (b) Experimental (stars) and simulated (circles) admittance spectrum for the frequency $f$ = 205 kHz. (c–d) The simulated voltage distributions for the $n$ = 2 Landau modes under varying pseudo-electric fields and non-reciprocal parameters. (c) The Landau modes under different values of $β$. (d) The Landau modes under different values of $δC$. (e–h) Theoretical simulations and (i–m)

experimental results of voltage distributions. (e, f) and (i, j) correspond to the first-order Landau level ($n = 1$), while (g, h) and (k, l) correspond to the second-order Landau level ($n = 2$).